# A Concise Survey of G4U

Dong Li[1, 2 †], Yunhua Zhang[1, 2, 3], and Liting Liang[1, 2, 3]

[1] CAS Key Laboratory of Microwave Remote Sensing

[2] National Space Science Center, Chinese Academy of Sciences

[3] University of Chinese Academy of Sciences

[†] lidong@mirslab.cn

October 31, 2019

## ABSTRACT

The general four-component model-based decomposition with unitary transformation of coherency matrix (G4U) is a state-of-the-art four-component decomposition, which has received extensive attentions recently. A literature survey is carried out to indicate the overall influence, improvement, development, evaluation, and application of G4U. Totally, 137 literatures are found mentioning G4U in Google Scholar© until October 7, 2019, which can be attributed into 4 categories in terms of the degree of concentration and 17 subcategories according to the focus of attention. Among these literatures, 61 of them simply mention G4U mainly because it is a new four-component model-based decomposition, a typical model-based decomposition, or even a target decomposition. There are also 9 literatures which improve G4U and develop G4U-like decompositions with unitary transformation of coherency matrix. 20 literatures generally use G4U as a typical target decomposition for comparison or a pseudo-color visualization technique to show the performance of some developed approaches. There are also 47 literatures dedicated to critically evaluate and deeply apply G4U in the remote sensing of forestry, agriculture, wetland, snow, glaciated terrain, earth surface, manmade target, environment, and damages caused by earthquake, tsunami, and landside, which indicates the value and significance of G4U in the true sense.

# I. INTRODUCTION

This paper is dedicated to contribute a review on the general four-component scattering power decomposition with unitary transformation of coherency matrix (G4U). G4U was developed by Yamaguchi *et al.* and Singh *et al.* in the following three papers:

[G4U1] G. Singh, Y. Yamaguchi, and S.-E. Park, "General four-component scattering power decomposition with unitary transformation of coherency matrix," *IEEE Transactions on Geoscience and Remote Sensing*, vol. 51, no. 5, pp. 3014-3022, May 2013.

[G4U2] Y. Yamaguchi, G. Singh, S.-E. Park, and H. Yamada, "Scattering power decomposition using fully polarimetric information," in *Proceedings of IEEE International Geoscience and Remote Sensing Symposium (IGARSS)*, Munich, Germany, 2012, pp. 91-94.

[G4U3] G. Singh, Y. Yamaguchi, Y. Cui, S.-E. Park, and R. Sato, "New four component scattering power decomposition method," in *Proceedings of European Conference on Synthetic Aperture Radar (EUSAR)*, Nuremberg, Germany, 2012, pp. 521-522.

The four-component model-based scattering power decompositions, like Yamaguchi's original four-component model-based decomposition (Y4O), Y4O with rotation of the coherency matrix (Y4R), and Y4R with extended volume scattering model (S4R) cannot provide a full utilization of all the nine polarimetric parameters contained in coherency matrix. To solve this problem, G4U introduces an additional unitary transformation in Y4R and S4R. As a result, G4U is claimed to make full use of the polarimetric scattering information, and is identified as the state-of-the-art four-component decomposition, which has received extensive attentions recently.

# II. METHOD

To reflect the overall citation, influence, improvement, development, evaluation, and applications of G4U so far, we conduct a concise survey by investigating the citation of the above three G4U literatures in Google Scholar© (http://scholar.google.com/). We choose Google Scholar because it includes not only literatures of multiple forms, such as papers, books, reports, and dissertations, but also literatures presented in multiple languages. The survey is time-consuming because we have to examine each candidate literature for pertinence, but it can obtain a comprehensive result. We believe this will be helpful to the polarimetric synthetic aperture radar (PolSAR) remote sensing area.

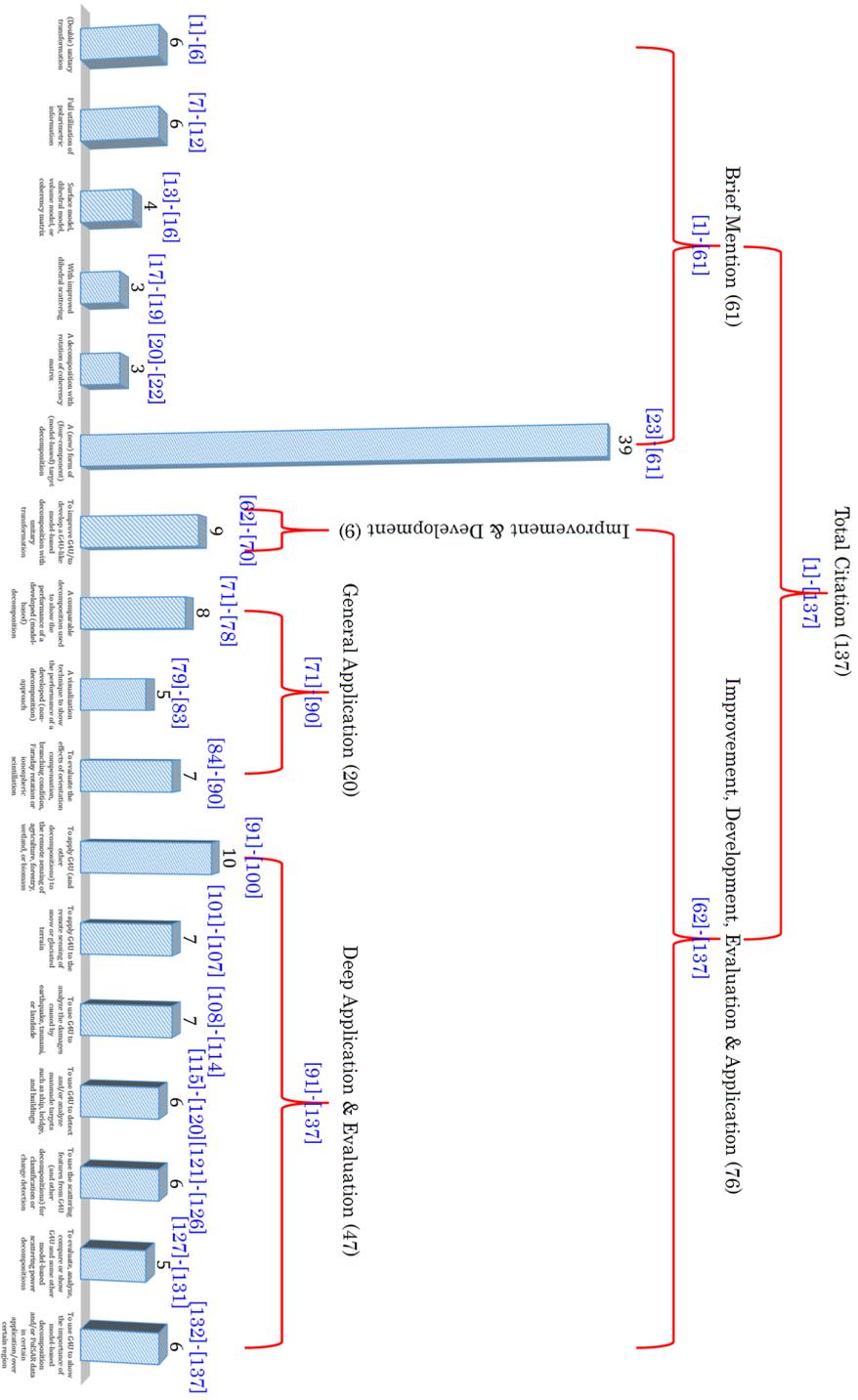

Fig. 1. Literature survey result of G4U. The survey is carried out by investigating the citation (until October 7, 2019) of the three G4U literatures [G4U1]-[G4U3] shown in Section I in Google Scholar©. The involved 137 literatures can be further attributed into 4 categories and 17 subcategories according to their focus of attention and degree of concentration. The number of involved literatures in each category and subcategory are shown in the parentheses, while the literatures are indexed by their order in References.

## III. RESULTS

Totally, 137 literatures are found citing [G4U1]-[G4U3] in Google Scholar until October 7, 2019, which are summarized in the References. The majority of them are presented in English (128), but there are also some literatures presented in Chinese (5), Japanese (2), German (1), and Russian (1). According to the focus of attention and the degree of concentration, we can attribute the literatures into 4 categories and 17 subcategories, as shown in Fig. 1.

● **Category 1: Brief Mention [1]-[61] (61÷137=44.52%)**
In Category 1, G4U is only simply mentioned without any improvement, development, or application. This involves in 61 literatures [1]-[61], which can be further attributed into six subcategories based on their focuses of attention.

✓ **Subcategory 1: (Double) unitary transformation [1]-[6] (6÷137=4.38%)**
In this subcategory, G4U is simply mentioned because it serially adopts a real rotation and an imagery rotation to achieve a double unitary transformation of coherency matrix so as to eliminate the number of observation parameters. Particularly, a new SU(3) matrix was formulated in the three literatures of G4U to achieve the imagery rotation. The physical significance of this new unitary transformation and its application also attract certain attention.

✓ **Subcategory 2: Full utilization of polarimetric information [7]-[12] (6÷137= 4.38%)**
In this subcategory, G4U is simply mentioned because the nine polarimetric scattering parameters of coherency matrix are all used in the four-component scattering models of G4U due to the use of the double unitary transformation. Thus, G4U is identified as a state-of-the-art four-component scattering power decomposition and has received extensive attentions recently.

✓ **Subcategory 3: Surface model, dihedral model, volume model, or coherency matrix [13]-[16] (4÷137=2.92%)**
In view of the fact that G4U has been accepted as a state-of-the-art four-component scattering power decomposition, G4U is even cited in this subcategory in the presentation of the surface scattering model, dihedral scattering model, volume scattering model, and even the coherency matrix. Particularly, like S4R, G4U also adopts an extended volume scattering model to distinguish volume scattering between dipole and dihedral scattering structures caused by cross-polarized HV component, which also draws certain attention.

✓ **Subcategory 4: With improved dihedral scattering [17]-[19] (3÷137=2.19%)**
The literature [G4U1] listed above indicated that G4U could enhance the double-bounce scattering contributions over urban area compared with the four-component decomposition such as Y4R and S4R because of the full utilization of polarimetric information. This makes G4U more suitable for the extraction of the built-up urban area which not only attracts the attention of literatures [17]-[19] but also lays the foundation for the application of G4U to assess and monitor the damages caused by natural disasters such as tsunami and earthquake [108]-[114]. We will investigate the latter later in Subcategory 13.

✓ **Subcategory 5: A decomposition with rotation of coherency matrix [20]-[22] (3÷137=2.19%)**

Rotation of the coherency matrix in target decomposition was first developed to improve Y4O to Y4R. It has been one important and essential preprocessing procedure for model-based target decomposition. As a state-of-the-art four-component scattering power decomposition, G4U is simply mentioned in this subcategory [20]-[22] because of the rotation of coherency matrix.

✓ **Subcategory 6: A (new) form of (four-component) (model-based) target decomposition [23]-[61] (39÷137=28.46%)**

In this subcategory, G4U is simply cited because it is a latest or typical version of the four-component model-based decomposition. In view of the important role played by four-component model-based decomposition, G4U is also cited in Subcategory 6 as a typical model-based decomposition or even as a typical target decomposition. Subcategory 6 contains the most 39 literatures among all the 17 subcategories. This indicates the wide recognition of G4U in the field.

● **Category 2: Improvement & Development [62]-[70] (9÷137=6.57%)**

In Category 2, G4U is further improved and some other G4U-like model-based decompositions with unitary transformation of coherency matrix are also developed. Totally, this involves in 9 literatures [62]-[70] only and we wholly reserve them in Subcategory 7 with no further distinction.

✓ **Subcategory 7: Improvement and development of G4U [62]-[70] (9÷137=6.57%)**

The idea of unitarily transforming the coherency matrix employed in G4U is followed in this subcategory to improve G4U and to contribute some G4U-like model-based target decompositions. Some other SU(3) matrices are involved in literatures [62]-[70] as an alternative or compensation to that used in G4U.

● **Category 3: General Application [71]-[90] (20÷137=14.60%)**

In Category 3, G4U is generally used by some authors as a typical target decomposition for comparison or as a pseudo-color visualization technique to show the performance of some other proposed algorithms. This involves in 20 literatures [71]-[90], which can be further attributed into three subcategories according to their focuses of attention.

✓ **Subcategory 8: A comparable decomposition used to show the performance of a developed (model-based) decomposition [71]-[78] (8÷137= 5.84%)**

In this subcategory, G4U is treated as a typical comparative decomposition to illustrate the performance of some other decompositions, such as the six- and seven-component scattering power decomposition [72], [73], [77], hybrid-pol decomposition [75], multi-component decomposition [76], three-component decomposition with reflection symmetry approximation [78], and two other model-based decompositions [71], [74].

✓ **Subcategory 9: A visualization technique to show the performance of a developed (non-decomposition) approach [79]-[83] (5÷137= 3.65%)**

In this subcategory, G4U is used as a pseudo-color visualization technique to display a PolSAR data [79] or a data preprocessed by another developed (non-decomposition) approach such as the phase error compensation [80] and the hybrid-pol reconstruction [81]-[83]. The visualization is achieved by assigning the powers of double-bounce scattering, volume scattering, and surface scattering decomposed by G4U as the color components of red, green, and blue, respectively.

✓ **Subcategory 10: To evaluate the effect of orientation compensation, branching condition, Faraday rotation or ionospheric scintillation [84]-[90] (7÷137 = 5.11%)**

Target orientation will influence the correct interpretation and understanding of the polarimetric scattering. The effect of deorientation on target extraction [84], [89], and the benefit of PolSAR data arrangement with rotation transformation [90] are evaluated and indicated by decomposing with G4U. Branching condition is an important procedure for four-component model-based target decomposition. Some different branch conditions are used in decompositions. Their effect on decomposition is also evaluated by G4U decomposing the data with different branch conditions [88]. As for a spaceborne PolSAR working in low frequency, Faraday rotation and ionospheric scintillation will also impact the acquired PolSAR image. The influence of ionospheric scintillation [86] and the performance of the developed Faraday rotation correction and mitigation approaches [85], [87] are also evaluated by decomposing the data with G4U.

● **Category 4: Deep Application and Evaluation [91]-[137] (47÷137=34.31%)**

In Category 4, G4U is deeply evaluated and applied. This indicates the importance and value of G4U in the true sense. This involves in 47 literatures [91]-[137], which can be further attributed into seven subcategories according to their focuses of attention.

✓ **Subcategory 11: To apply G4U (and other decompositions) to the remote sensing of agriculture, forestry, wetland, or biomass [91]-[100] (10÷137=**

7.30%)

G4U and/or some other target decompositions are applied in this subcategory to processing PolSAR datasets for the remote sensing of agriculture (such as rice phenology [94]), forestry (such as mangroves [91], deciduous forest [92], tropical acacia plantation [93], [97], and rubber trees [99]), wetland [96], land cover [98], and the aboveground biomass [95], [100].

✓ **Subcategory 12: To apply G4U to the remote sensing of snow or glaciated terrain [101]-[107] (7÷137=5.11%)**

G4U is utilized in this subcategory to estimate snow density [101] and wetness [103]-[106], to map the dry/wet snow [107], and to categorize the glaciated terrain [102]. The literature [G4U1] listed in Section I indicated that G4U could also enhance surface scattering power over areas where surface scattering is dominant, such as the snow and glaciated terrain. This lays the foundation for the application of G4U to the remote sensing of snow and glaciated terrain.

✓ **Subcategory 13: To use G4U to analyze the damages caused by earthquake, tsunami, or landside [108]-[114] (7÷137=5.11%)**

The literature [G4U1] shown above indicated that G4U could enhance surface scattering power over areas where surface scattering is dominant (such as the snow, glaciated terrain, and landslide) and enhance double-bounce scattering power over regions where double-bounce scattering is dominant (such as the buildings and urban area). Hence, G4U is used in this subcategory to monitor, map, extract, and evaluate the earthquake/tsunami damages by analyzing the double-bounce scattering power before and after disasters [108]-[110], [113], and to recognize and extract landslides and damaged bridge by comparing the surface scattering power before and after disasters [111], [112], [134].

✓ **Subcategory 14: To use G4U to detect and/or analyze manmade targets such as ship, bridge, and buildings [115]-[120] (6÷137=4.38%)**

In this subcategory, the polarimetric scattering features of target extracted by G4U, mainly the surface scattering power and the double-bounce scattering power, are used to detect ship [115], [116], [119], to extract manmade targets [118], and to analyze backscattering characteristics of bridges [117] and buildings [120].

✓ **Subcategory 15: To use the scattering features from G4U (and other decompositions) for classification or change detection [121]-[126] (6÷137=4.38%)**

In this subcategory, the polarimetric scattering features of target extracted by G4U and/or some other decompositions such as entropy/alpha are integrated to obtain an unsupervised or optimal classification of PolSAR data [121]-[123], [125], [126], and to enable a G4U-based change detector [124].

✓ **Subcategory 16: To evaluate, analyze, compare or show G4U and some other model-based scattering power decompositions [127]-[131] (5÷137=3.65%)**

The widely-used model-based scattering power decompositions, particularly, the four-component model-based decompositions like G4U are quantitatively evaluated [127], critically analyzed [128], [129], and qualitatively investigated [130], [131] in this subcategory.

**Subcategory 17: To use G4U (and other decompositions) to show the importance of model-based decomposition and/or PolSAR data (acquired by certain spaceborne/airborne system) in certain application/over certain region [132]-[137] (6÷137=4.38%)**

In this subcategory, the importance of PolSAR and model-based decomposition, particularly, the significance of the scattering powers is demonstrated by decomposing airborne (such as Pi-SAR2 [132]) and spaceborne PolSAR images acquired by ALOS-PALSAR2 [133] and ALOS-PALSAR1 [137] over areas such as Amazon River [134], Taiwan [135], and Singapore [137] with G4U and/or some other model-based decompositions for environment [134] and earth surface monitoring [136].

## IV.  CONCLUSION

G4U is the state-of-the-art four-component decomposition and has received extensive attentions recently. The full utilization of polarimetric information makes G4U a typical form of (four-component) (model-based) target decomposition which draws the most citation among all the 4 categories and 17 subcategories, and also makes G4U a typical comparative decomposition or a nice pseudo-color visualization technique to indicate the performance of other decompositions and algorithms. The successful attempt and deep application of G4U for the remote sensing of forestry, agriculture, wetland, earth surface, snow, glaciated terrain, manmade target, environment, and damages caused by earthquake, tsunami, and landside are possible mainly because G4U could enhance surface scattering power over area where surface scattering is dominant, and enhance double-bounce scattering power over region where double-bounce scattering is dominant, which indicates the value and significance of G4U in the true sense. We believe this concise survey will be beneficial to the PolSAR remote sensing area.

# REFERENCES

--------------------------------------------------------------------------------------------------------

- **Brief Mention**

--------------------------------------------------------------------------------------------------------

✓ **(Double) unitary transformation**

--------------------------------------------------------------------------------------------------------

--------------------------------------------------------------------------------------------------------

✓ **Full utilization of polarimetric information**

--------------------------------------------------------------------------------------------------------

-----------------------------------------------------------------------------------------------------

✓ **Surface model, dihedral model, volume model, or coherency matrix**

-----------------------------------------------------------------------------------------------------

-----------------------------------------------------------------------------------------------------

✓ **With improved dihedral scattering**

-----------------------------------------------------------------------------------------------------

-----------------------------------------------------------------------------------------------------

✓ **A decomposition with rotation of coherency matrix**

-----------------------------------------------------------------------------------------------------

52, no. 3, pp. 1705-1718, 2013.

---

✓ **A (new) form of (four-component) (model-based) target decomposition**

---

-------------------------------------------------------------------------------------------------------

- **Improvement & Development**

-------------------------------------------------------------------------------------------------------

------------------------------------------------------------------------------------------------

## ● General Application

------------------------------------------------------------------------------------------------

✓ **A comparable decomposition used to show the performance of a developed (model-based) decomposition**

------------------------------------------------------------------------------------------------

---

✓ **A visualization technique to show the performance of a developed (non-decomposition) approach**

---

---

✓ **To evaluate the effects of orientation compensation, branching condition, Faraday rotation or ionospheric scintillation**

---

-----------------------------------------------------------------------------------------------------------

● **Deep Application & Evaluation**

-----------------------------------------------------------------------------------------------------------

✓ **To apply G4U (and other decompositions) to the remote sensing of agriculture, forestry, wetland, or biomass**

-----------------------------------------------------------------------------------------------------------

-------------------------------------------------------------------------------------------------------

✓ **To use G4U to detect and/or analyze manmade targets such as ship, bridge, and buildings**

-------------------------------------------------------------------------------------------------------

-------------------------------------------------------------------------------------------------------

✓ **To use the scattering features from G4U (and other decompositions) for classification or change detection**

-------------------------------------------------------------------------------------------------------

-------------------------------------------------------------------------------------------------

✓ **To evaluate, analyze, compare or show G4U and some other model-based scattering power decompositions**

-------------------------------------------------------------------------------------------------

-------------------------------------------------------------------------------------------------

✓ **To use G4U (and other decompositions) to show the importance of model-based decomposition and/or PolSAR data (acquired by certain spaceborne/airborne system) in certain application/over certain region**

-------------------------------------------------------------------------------------------------

-----------------------------------------------------------------------------------------------------------------